\documentclass[aps,prb,preprint,showpacs,superscriptaddress]{revtex4-1}
\usepackage{graphicx}
\usepackage{bm}

\begin{document}

%\preprint{}

\title{Equilibrium rotation of a vortex bundle terminating on a lateral wall}


\author{E. B. Sonin}
%\thanks{}
\affiliation{Racah Institute of Physics, Hebrew University of
Jerusalem, Jerusalem 91904, Israel}

\author{S. K. Nemirovskii}
\affiliation{Institute of Thermophysics, Prospect Lavrentyeva, 1, 630090, Novosibirsk,
Russia}


\date{\today}

\newcommand{\be}{\begin{equation}}
\newcommand{\ee}[1]{\label{#1}\end{equation}}
\newcommand{\bem}{\begin{eqnarray}}
\newcommand{\eem}[1]{\label{#1}\end{eqnarray}}
\newcommand{\eq}[1]{Eq.~(\ref{#1})}
\newcommand{\Eq}[1]{Equation~(\ref{#1})}

\begin{abstract}
The paper investigates possibility of equilibrium solid-body rotation of a vortex bundle
diverging at some height from a cylinder axis and terminating on a lateral wall of a container.
Such a bundle arises when vorticity expands up from a container bottom  eventually filling
the whole container. The analysis starts from a single vortex, then goes to a vortex sheet,
and finally addresses a multi-layered crystal vortex bundle. The equilibrium solid-body
rotation of the vortex bundle requires that the  thermodynamic potentials in the vortex-filled
and in the vortex-free parts of the container are equal providing the absence of a force
on the vortex front separating the two  parts. The paper considers also a weakly
non-equilibrium state when the bundle and the container rotate with different angular
velocities and the vortex front propagates with the velocity determined by 
friction between vortices  and  the container or the normal liquid moving together with the container.
\end{abstract}

\pacs{67.30.hb,47.15.ki, 67.30.he}

%\keywords{}

\maketitle

%\section{}
%\subsection{}
%\subsubsection{}

% If in two-column mode, this environment will change to single-column
% format so that long equations can be displayed. Use
%\begin{widetext}
%\end{widetext}

% figures should be put into the text as floats.
% Use the graphics or graphicx packages (distributed with LaTeX2e)
% and the \includegraphics macro defined in those packages.
% See the LaTeX Graphics Companion by Michel Goosens, Sebastian Rahtz,
% entire page. There is no need to do explicit centering.

\section{Introduction}

Transient processes of establishing of stable vorticity in rotating
superfluids have always been in the focus of attention of theorists and experimentalists
studying superfluid vortex dynamics. An important example of such a process
is penetration of a vortex bundle into an originally vortex-free rotating
container filled with a superfluid. This process was thoroughly investigated
in superfluid $^3$He-B theoretically and experimentally \cite%
{Twist,Lamin,Kru}. Vorticity is generated at the container bottom and
propagates upward along the cylindric container axis in the form of a vortex
bundle flaring to lateral container walls. The flaring part of the vortex
bundle was called \emph{vortex front}. Below the front the vortex bundle is
vertical but twisted. The twist is connected with the flux of the angular
momentum along the bundle, which must dissipate due to either mutual
friction in the bulk or friction at rough wall surface. A great attention
was directed to transition from laminar to turbulent vortex front
propagation, especially at low temperature where disappearance of mutual
friction facilitates the transition to turbulence.

All studies of the flaring vortex bundle, analytical, numerical, and
experimental, were performed in the presence of dissipation, without which
front propagation is impossible since it is accompanied by change of the
total energy and the angular momentum. The goal of the present work was to
check whether a stable solution for a vortex bundle terminating on a lateral
wall may exist as an equilibrium state without dissipation and propagation
along the rotation axis. In this equilibrium state the whole bundle together
with its vortex front rotates without twisting as a solid with constant angular
velocity. If the container rotates with the same angular velocity neither
dissipation nor propagation of the vortex front along the rotation axis is
possible. Our paper considers conditions for existence of such
``eigenrotation'' and analyses the effect of weak friction, which
leads to propagation of the vortex front if the container and the bundle
rotate with different angular velocities.

The paper starts from the predecessor of the vortex bundle: a single vortex filament, 
located on the container axis in the lower part of the container, at its higher part continuously goes away from the axis
and eventually terminates on the lateral wall (Sec. \ref{SinVor}). Section \ref{SinShe}  analyses
a vortex bundle in which vortices form a single axisymmetric layer (vortex sheet).
Section \ref{twoShe} addresses the case of two coaxial non-interacting sheets and
demonstrates that it is impossible to find the equilibrium solid-body rotation for 
such a two-layer bundle. This led to conclusion that equilibrium solid-body rotation
of a multi-layered vortex bundle requires effective interaction between layers.
This could be the same interaction, which leads to formation of a vortex crystal.
The condition of equilibrium rotation of a stiff solid-body vortex bundle diverging
to a lateral wall at some height is analyzed in Sec. \ref{solidBun}.
Section \ref{fric} discusses weakly non-equilibrium vortex bundle, when 
friction of the rotating vortex bundle makes possible propagation of the vortex front along the
rotation axis. Concluding remarks are presented in the last Sec. \ref{concl}.

\section{Single vortex} \label{SinVor}

\begin{figure}[tbp]
\includegraphics[scale=0.55]{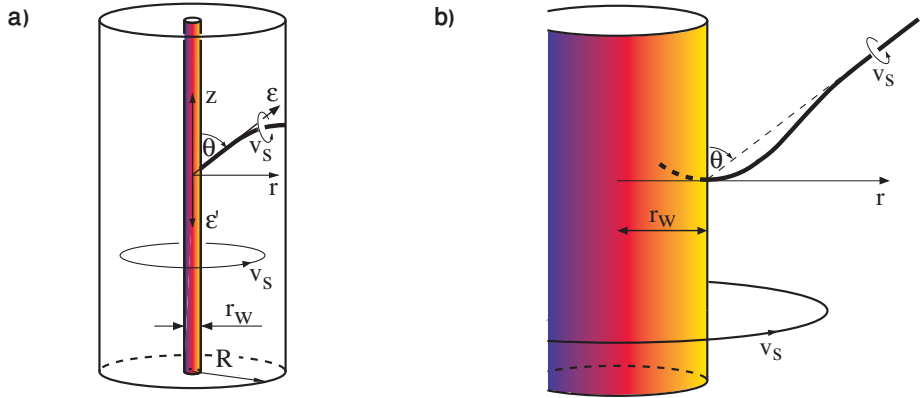}
% note that the square brace option below is only required % if you intend to produce a list of illustrations
\caption{(Color online) Vortex line attached to a thin wire.The profile of the vortex line can be examined on two different
length scales: (a) At large scales $r\gg r_w$ the local induction
approximation becomes applicable, in which the wire is treated as an
enhanced vortex core. The vortex line meets a wire at the finite contact
angle $\protect\theta$. (b) On small scales $r \sim r_w$ the vortex line
goes smoothly over into the image vortex (dashed continuation of the vortex
line), with the wire surface perpendicular at the connection point.}
\label{f4-3}
\end{figure}

Let us start from the most elementary case when the bundle reduces to a
single vortex, which terminates on a lateral wall of a container. A freely
precessing vortex is in fact a particular case of the geometry, which  has
already been carefully investigated theoretically and experimentally in
connection with the study of precession of single vortex trapped on a wire
coaxial with a cylindric container filled by superfluid $^3$He-B \cite%
{Zieve,MV,Schw,Curv,SK}. Geometry of the experiment \cite{Zieve} is shown in
Fig.~\ref{f4-3}(a). The $z$-axis is the axis of the cell of the radius $R$
and of the wire of the radius $r_w$. The vortex filament with the
circulation $\kappa$ of the superfluid velocity being trapped on the wire at
$z<z_0$, peels from the wire at $z=z_0$ stretching to the container lateral
wall. Just the ``unzipped'' (free) part of the vortex filament at $z>z_0$
participates in the precession.

If dissipation is absent, the vortex-line shape may be found by minimization
of the energy in the coordinate frame rotating with the angular velocity $
\Omega$ of the vortex precession. This energy corresponds to the Gibbs
thermodynamic potential
\begin{equation}
G =E- \Omega M_z , \label{rot}
\end{equation}
where
\begin{eqnarray}
E=z_0 {\frac{\rho_s \kappa ^2}{4\pi}} \ln{\frac{R}{r_w}} + +
\int\limits_{r_w}^{R} {\frac{\rho_s \kappa ^2}{4\pi}} \ln{\frac{R}{r_c}}
\sqrt{1 + \left[ \frac{dz(r)}{dr} \right]^2}\, dr \label{energy}
\end{eqnarray}
is the energy in the local induction approximation,
\begin{equation}
M_z={\frac{\rho_s\kappa}{2}}(R^2-r_w^2) z_0 + \rho_s \kappa \int_{r_w}^{R}[
z(r)-z_0] r dr  \label{moment}
\end{equation}
is the angular momentum of the liquid around the axis $z$, $r_c$ is the vortex core
radius, and the shape of the vortex line in the
cylindric coordinates is given by the function $z(r)$. It is supposed that
the line is inside the axial plane, so the azimuthal angle $\phi$ does not
vary along the line. The function $z(r)$ is determined by variation of the
thermodynamic potential with respect to the vertical coordinate $z$ of the
vortex line. The corresponding Euler--Lagrange equation is
\begin{equation}
\rho_s \kappa \Omega r = - \frac{d}{dr} \left[ \frac{\varepsilon (r)\,dz/dr}{%
\sqrt{1+(dz/dr)^2}}\right] ,  \label{EL}
\end{equation}
where
\be
\varepsilon = \frac{\rho_s \kappa ^2}{4\pi} \ln{\frac{R}{r_c}}
    \ee{LT}
is the energy per unit length of the vortex line, which determines the line-tension force.
\Eq{EL} was derived from minimization of the thermodynamic
potential, but  at the same time it is the equation of balance of forces on the precessing
vortex line: the Magnus force on the left-hand side, which is determined by
the linear velocity $\Omega r$ of the precession, is balanced by the line
tension force $\propto \varepsilon$.
One should look for a solution of this equation with the boundary conditions
on the lateral wall and the wire. The first one is the condition of
transversality (the vortex line normally ends at the lateral wall of the
cylinder):
\begin{equation}
\left. \frac{dz}{dr}\right\vert _{r=R}=0.  \label{transvers}
\end{equation}%
The second boundary condition is imposed in the ``unzipping'' point $z=z_0$:
\begin{equation}
\cos\theta =\left.\left[\frac{dz/dr}{ \sqrt{1+(dz/dr)^2}}\right]%
\right|_{r=r_w}= \frac{\varepsilon}{\varepsilon_w} = \frac{\ln{\frac{R}{r_w}}%
}{\ln{\frac{R}{r_c}}},  \label{5}
\end{equation}
where $\theta$ is the angle between the $z$ axis and the vortex line at
$z=z_0$ and $\varepsilon_w =(\rho_s\kappa^2/4\pi) \ln{(R/r_w)}$ is the line
tension of the vortex filament attached to the wire.
The condition  directly
follows from minimization of the Gibbs potential with respect to variation
of the unzipping coordinate $z_0$ \cite{Curv}.  It ensures the balance  of line-tension forces along the $z$ axis for the
element of the vortex filament in the  point $z=z_0$ (see Fig.~\ref{f4-3}):
The unzipping point is at rest if $\varepsilon_w=\varepsilon \cos \theta$.  In our approach the trapped segment
of the vortex line is treated as a vortex line with a larger core radius $r_w$.
So the approach is valid only for rather thin wires with radius $r_w$ much
less than the radius $R$ of the container.

Integrating Eq.~(\ref{EL}) over $r$ with the transversality condition Eq.~(%
\ref{transvers}) one obtains the first integral
\begin{equation}
\frac{\rho_s \kappa}{2} \Omega(r^2- r_w^2)=\frac{\rho_s \kappa^2}{4\pi}\ln{%
\frac{R}{r_c}}\left[\frac{dz/dr}{ \sqrt{1+(dz/dr)^2}}\right].  \label{1int}
\end{equation}
At the unzipping point this equation together with boundary condition Eq.~(%
\ref{5}) yields that
\begin{equation}
g= e - \Omega m_z =0,  \label{eqG}
\end{equation}
where $g$, $e= \varepsilon_w $, and $m_z= \rho_s \kappa \Omega(r^2- r_w^2)/2$ are the Gibbs
potential, the energy and the angular momentum per unit length of the vortex line
below the unzipping point. Emergence of the condition of zero Gibbs potential
below the unzipping point from the minimization of the total Gibbs potential
is quite natural. In the vortex-free region the Gibbs potential density
vanishes, and if the Gibbs potential density below the unzipping point is
nonzero, then there is a force driving the curved piece of the vortex line
(an analog of the vortex front) upward or downward. The condition (\ref{eqG}%
) determines the precession angular velocity of the curved vortex line
stretched between the wire and the cell lateral wall \cite{Zieve}:
\begin{equation}
\Omega=\frac{\kappa}{2\pi(R^2-r_w^2)} \ln{\frac{R}{r_w}}\approx \frac{\kappa
}{2\pi R^2} \ln{\frac{R}{r_w}}.  \label{4.01}
\end{equation}
Note that this expression is exact and does not depend on vortex line shape
of the curved segment or on using the local induction approximation \cite%
{Zieve,MV,Curv,Schw}. This is a direct consequence of the rigorous canonic
relation
\begin{equation}
\Omega =\frac{\partial E}{\partial M_z}.  \label{kinrot}
\end{equation}
Varying the position of the unzipping point the variations of the energy and
momentum are $dE =\varepsilon_wdz_0$ and $dM_z =m_z dz_0$, so $\partial
E/\partial M_z =\varepsilon_w/m_z$, and Eq.~(\ref{kinrot}) yields expression Eq.~(\ref{4.01}) for the precession angular frequency.

This analysis can be applied to the case of a free axial vortex ending at
the wall. The case is the limit of an extremely thin wire when the wire
radius must be replaced by the vortex core radius in all expression.
Formally this means that the unzipping point goes to $z \to -\infty$, and
the curved vortex line smoothly approaches to the vertical axis of rotation
in accordance with the boundary condition Eq.~(\ref{5}), which now tells
that
\begin{equation}
\left.{\frac{dz}{dr}} \right|_{r =0}\to \infty.
\end{equation}
The shape of the free vortex can be analytically obtained after the second
integration of Eq.~(\ref{1int}) taking into account the expression Eq.~(\ref%
{4.01}) for the angular velocity $\Omega$:
\begin{equation}
z(r)=\sqrt{2R^2 -r^2}-R-{\frac{R}{\sqrt{2}}}\ln \frac{R\sqrt{2}+\sqrt{2R^2
-r^2}}{r(\sqrt{2}+1)}.
\end{equation}
The expression demonstrates that the vortex line exponentially approaches to
the axis $r=0$: $r \approx R e^{-|z | \sqrt{2}/R}$.

While the frequency of the single vortex precession straightforwardly
follows from commonly accepted thermodynamic arguments, the shape of the vortex line was a
matter of dispute resulting from disagreement on a proper usage of the local
induction approximation for the precessing partially trapped vortex line.
One may find a detailed discussion of the issue in Refs. \onlinecite{Curv,SK}. In particular, the debate was about a proper boundary condition at the unzipping
point. Instead of the boundary condition Eq.~(\ref{5}) based on the balance
of forces directly following from the variational principle Schwarz\cite{Schw}
used the condition that the vortex line is normal to the wire. At the very
surface of the wire the latter boundary condition is definitely correct. But
at small scales of the order of the wire radius $r_w$ there are forces,
which led to fast deviation from the normal direction. These forces were
neglected by Schwarz and all others addressing this problem. This is a
legitimate approximation when one looks for the vortex shape at large scales
of the order of the container radius $R$ but only if one uses the boundary
condition (\ref{5}) based on the balance of line-tension forces. In reality
this means that the boundary condition is imposed not exactly at the radius $r_w$
of the wire but on the distance larger than $r_w$, which at the same time is
still much smaller than $R$, as illustrated in Fig.~\ref{f4-3}b. It is
worthwhile to note that the analysis of the shape of a free precessing
vortex gives one more justification of the force-balance boundary condition.
The latter provides a natural transition from the vortex partially trapped
by the wire to a free vortex smoothly changing its direction from vertical to
horizontal. On the other hand, Schwarz's boundary condition
becomes senseless for the free vortex since it requires that the vortex
meets the axis normally.

\section{Single vortex sheet} \label{SinShe}

Our next step is to analyze a bundle of vortices but still of simple
geometry: $N_1 $ vortices form a cylindric vertical sheet of radius $r_1$
which at some height diverges to lateral wall forming a whorl (see Fig. 2). The
single-sheet whorl is a simulation of a more complicated vortex front. As in
the case of a single vortex, any vortex line in the sheet is given by the
function $z(r)$, which is independent on $\phi$ and is determined from the
variational principle for the the total Gibbs potential $G=E-\Omega M_z$. The
total angular momentum is
\begin{equation}
M_z={\rho }_{s}\int_{r_1}^R2\pi rdr \int^{z(r)}dz [v_s(r)r]={\rho }_{s}N_1\kappa
\int_{r_{1}}^{R}z(r)r\,dr,  \label{M1 cg}
\end{equation}%
where $v_s(r)=N_1\kappa /2\pi r$ is the azimuthal superfluid velocity induced by the
vortex sheet. The total kinetic energy $E=E_s+E_v$ consists of the energy
\begin{equation}
E_s={\rho }_{s}\int_{r_1}^R2\pi rdr \int^{z(r)}dz\frac{v_s(r)^2}{2}=\frac{{%
\rho }_{s}(N_1\kappa )^{2}}{4\pi }\int_{r_{1}}^{R} z(r)\frac{dr}{r}
\label{E1 cg}
\end{equation}%
of the velocity field induced by the vortex sheet and the energy of
individual vortex lines given by [compare with the expression Eq.~(\ref%
{energy}) for a single vortex]:
\begin{equation}
E_v=\frac{{\rho }_{s}N_1\kappa ^{2}}{4\pi }\int_{r_{1}}^{R}\ln {\frac{b}{r_c}%
}\sqrt{1+\left[\frac{dz(r)}{dr}\right]^{2}}dr ,  \label{E1 vort}
\end{equation}%
where $b=2\pi r/N_1$ the $r$-dependent intervortex spacing.

\begin{figure}[tbp]
\includegraphics[width=0.4\linewidth]{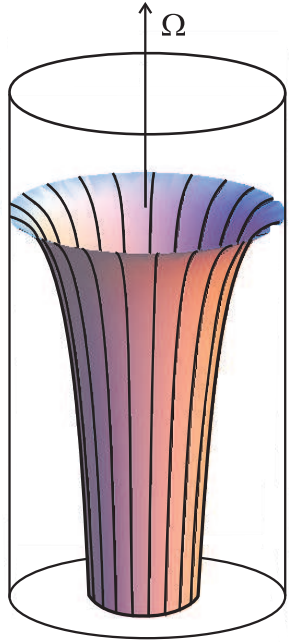}
\caption{(Color online) Single vortex sheet diverging to lateral walls via  forming a whorl}
\label{fig2}
\end{figure}

Variation of the Gibbs potential of the vortex sheet yields the
Euler--Lagrange equation:
\begin{equation}
{\rho }_{s}N_1\kappa \Omega r-\frac{{\rho }_{s}(N_1\kappa )^{2}}{4\pi }\frac{%
1}{r}=\frac{N_1{\rho }_{s}\kappa ^{2}}{4\pi }\frac{d}{dr}\ln \left( \frac{%
2\pi r}{N_1r_c}\right) \frac{dz/dr}{\sqrt{1+(dz/dr)^{2}}}.
\label{Euler eq 1}
\end{equation}%
The first integral of this equation for the boundary condition Eq.~(\ref%
{transvers}) at the container lateral wall is
\begin{equation}
\frac{{\rho }_{s}N_1\kappa \Omega }{2}\left( r^{2}-R^{2}\right) -\frac{{\rho
}_{s}(N_1\kappa )^{2}}{4\pi }\ln \left( \frac{r}{R}\right) +\frac{N{\rho }%
_{s}\kappa ^{2}}{4\pi }\ln \left( \frac{2\pi r}{N_1r_c }\right) \frac{dz/dr}{%
\sqrt{1+(dz/dr)^{2}}}=0.  \label{first integral 1,2}
\end{equation}%
The boundary condition
\begin{equation}
\left.{\frac{dz}{dr}} \right|_{r =r_1}\to \infty
\end{equation}
provides a transition from the whorl to the vertical stem of the vortex
sheet. Using it in Eq.~(\ref{first integral 1,2}) gives the condition that
the Gibbs potential per unit length $g_1=\varepsilon -\Omega m_z$ ($\varepsilon
$ and $m_z$ are the energy and the angular momentum per unit length of the stem)
vanishes at the stem:
\begin{equation}
g_1= \frac{{\rho }_{s}(N_1\kappa )^{2}}{4\pi }\ln \left( \frac{R}{r_1}%
\right) +\frac{N_1{\rho }_{s}\kappa ^{2}}{4\pi }\ln \left( \frac{2\pi r_1}{%
N_1r_c }\right) -\frac{{\rho }_{s}N_1\kappa \Omega }{2}\left(
R^{2}-r_1^{2}\right)=0.  \label{g1}
\end{equation}%
As in the case of a single vortex, this condition is necessary for the
absence of the force driving the front along the vertical axis.

But our minimization of the Gibbs potential requires an additional step:
minimization $g$ with respect to $r_{1}$ at fixed number $N$ of vortices.
The minimization yields that
\begin{equation}
\Omega =\frac{(N_{1}-1)\kappa }{4\pi r_{1}^{2}}.  \label{sold body 1}
\end{equation}%
The relation means that the whole sheet rotates as a solid body with linear
velocity $\Omega r_{1}=(N_{1}-1)\kappa /4\pi r_{1}$. Though our analysis was
based on the local induction approximation, for the vertical part (stem) of
the vortex sheet the obtained expression is exact and is easily derived from
the Bio--Savart law  for  $N_{1}$ equidistant vortices on the
circumference of the radius $r_{1}$. The modulus of the velocity induced by one vortex on another, $v_{ind} =\kappa/2\pi \delta$, is determined by the distance $\delta =2r_{1}\sin
(\alpha /2)$ between them, where $\alpha $ is the angle between radii directed from the
center to the both vortices. The azimuthal  component  $v_{ind}\sin (\alpha /2)=\kappa/4\pi r_1$ of the velocity does not depend on distance between two vortices, and therefore any vortex in the sheet moves along the circumference with the azimuthal velocity  $v_{sheet}= (N-1)\kappa/4\pi r_1$ induced by rest $N-1$ vortices. The radial component of any vortex vanishes for equidistant location of vortices by symmetry.  If $N_{1}\gg 1$ the velocity $v_{sheet}=1/2\left(
v_{in}+v_{out}\right) $ is an average of the velocities $v_{in}=0$ and $
v_{out}=N_{1}\kappa /2\pi r_{1}$ on both sides of the sheet (inside and outside). This law of motion for vortex sheets  is well known in classical hydrodynamics.

\begin{figure}[b]
\includegraphics[width=0.4\linewidth]{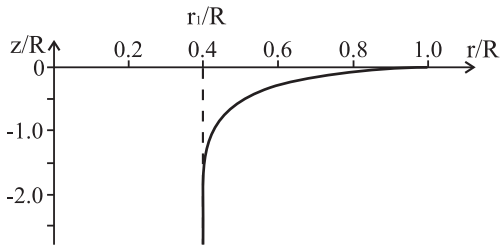}
\caption{The profile of the vortex sheet calculated from the Euler--Lagrange equation (\ref{Euler eq 1}) for $N_{1}=10$ and  $\Omega R^2/\kappa =4.52828$.}
\label{fig3}
\end{figure}

In Fig. \ref{fig3} we depicted the solution of the Euler--Lagrange equation (\ref{Euler eq 1}) with the boundary conditions $z(R)=0$ and $dz/dr\vert
_{R}=0$ for $N_{1}=10$ and  $\Omega R^2/\kappa =4.52828$. The whorl smoothly diverges from the vertical stem with the radius determined from Eq.~(\ref{sold body 1})
and terminates on the lateral wall.

\section{Two vortex sheets} \label{twoShe}

\begin{figure}[tbp]
\includegraphics[width=0.4\linewidth]{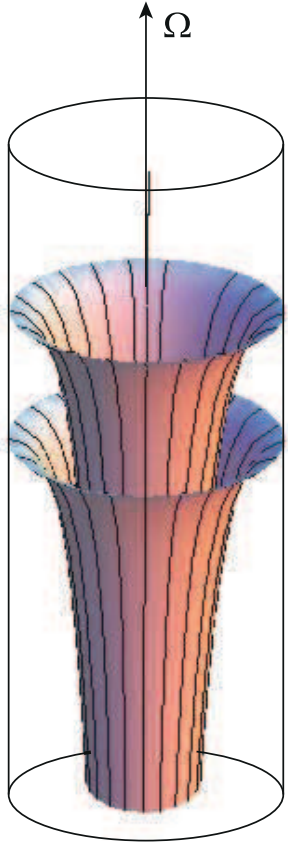}
\caption{(Color online) Hypothetical picture of two vortex sheets terminating on the lateral wall. But no state of solid-body rotation of two sheets with the same angular velocity was found.}
\label{f4}
\end{figure}

One might think that a more realistic vortex bundle could be modeled as an
ensemble of coaxial vortex sheets with their stems rotating together with
the same angular velocity as a solid body. In order to check this option we
considered two coaxial vortex sheets with numbers of vortices $
N_{1}$ and $N_{2}$ diverging to lateral walls via two whorls (Fig. \ref{f4}). The analysis of the Gibbs potential of the sheets  below the
whorls has shown that the condition for solid body rotation of the
two sheets with the same angular velocity cannot be realized. The density of the Gibbs potential for two vertical sheet is $g=g_1 +g_2$ where the Gibbs potential $g_1$ per unit length of the inner
sheet is given by Eq.~(\ref{g1}) and the Gibbs potential $g_2$ per unit
length of the outer sheet is
\begin{equation}
g_2= \frac{{\rho }_{s}\kappa ^{2}N_{2}(2N_{1}+N_{2})}{4\pi }\ln \left( \frac{%
R}{r_{2}}\right) +\frac{N_{2}{\rho }_{s}\kappa ^{2}}{4\pi }\ln \left( \frac{%
2\pi r_{2}}{N_{2}a_{0}}\right) -\Omega \frac{{\rho }_{s}N_{2}\kappa }{2}%
\left(R^{2}- r_{2}^{2}\right).  \label{F2 vert whorls}
\end{equation}%
Minimization with respect to $r_2$ yields the relation
\begin{equation}
\Omega =\frac{(N_2+2N_1-1)\kappa }{4\pi r_{2}^{2}} .  \label{sold body 2}
\end{equation}
The equilibrium condition (absence of the driving force on the whorl) is
satisfied only if $g_1=0$ and $g_2=0$. One may find the states with
vanishing $g_1$ and $g_2$ only if the angular velocities $\Omega$ in Eqs. (\ref{sold body 1}) and (%
\ref{sold body 2}) are  different. The state with the solid body rotation of
two sheets together cannot be found. However this outcome results from
shortcoming of our model and does not mean that the solid-body ``stem + whorl''
structure for multi-layer vortex bundles is impossible. In our model the
vortex sheet induces a fully axisymmetric velocity field outside the sheet
and no field inside. Though discrete vortex structure of the sheet is taken
into account via a logarithm contribution in the energy, the effect of
individual vortices on the velocity field is neglected. Such an
approximation is valid only if the distance between sheets exceeds the
intervortex distance in any sheet. In a tightly packed vortex bundle the
approximation fails, and interaction between neighboring layers glues them
effectively and does not allow rotation with different velocities. Then the
condition for an equilibrium vortex front becomes much less severe: it is
not necessary for the Gibbs potential of \emph{any } sheet to vanish, it is
sufficient that the \emph{total} Gibbs potential vanishes. The latter
condition will be considered in the next section.

\section{Equilibrium rotation of the solid-body vortex bundle terminating on the lateral wall} \label{solidBun}

From the analysis of a single and two vortex sheets it is evident that the
most restrictive condition for  equilibrium solid-body rotation of a vortex bundle terminating on the lateral wall is
zero Gibbs potential much below the front. If this condition is satisfied
the solution of the differential Euler--Lagrange equation for the front
shape is straightforward, though technically complicated: for a multilayered
vortex whorl one should solve a partial differential equation in the space
of two coordinates $r$ and $z$. This section addresses only the condition of
solvability of this equation: absence of a force on the vortex front.

\begin{figure}[tbp]
\includegraphics[width=0.2\linewidth]{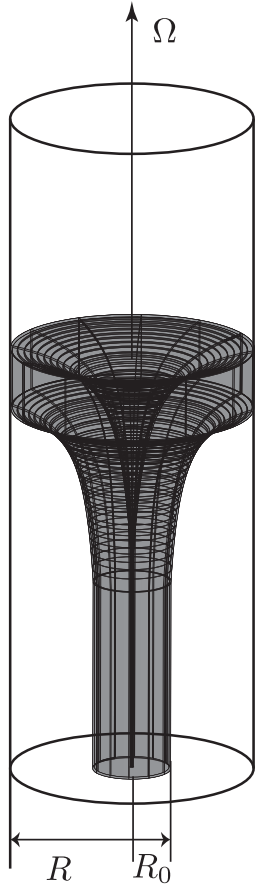}
\caption{(Color online) Solid-body vortex bundle terminating on the lateral wall.}
\label{f5}
\end{figure}

We look for the equilibrium state of fixed number $N$ of vortices forming the solid-body vortex bundle terminating on the lateral wall and rotating with the angular velocity $\Omega$ (Fig. \ref{f5}). 
The azimuthal velocity field in the stem of 
the bundle is
\begin{equation}
v=\left\{%
\begin{array}{cc}
\Omega r & r<R_0 \\
{\frac{\Omega R_0^2 }{r}} & r>R_0
\end{array}
\right. ,
\end{equation}
where $R_0$ is the radius of the bundle stem.
The number of vortices and the bundle radius are connected with the relation
$N=2\pi \Omega R_0^2/\kappa$. The Gibbs potential for the unit length of the
bundle stem in the coordinate frame rotating with the angular velocity $\Omega_0$
is
\begin{eqnarray}
g=\varepsilon-\Omega_0 m_z=\pi \rho_s \Omega^2 R_0^4\left( \ln{\frac{R}{R_0}} +%
{\frac{1 }{4}}\right)+{\frac{\rho_s\kappa \Omega R_0^2}{2}}\ln {\frac{r_v}{%
r_c}} -\Omega_0\pi \rho_s\Omega R_0^2\left( R^2-{\frac{R_0^2}{2}}\right).
\label{G0}
\end{eqnarray}
In the equilibrium $\Omega$ must coincide with $\Omega_0$ and using the
relation between $\Omega$ and $R_0$ this equation reduces to
\begin{eqnarray}
g={\frac{\rho_s\kappa^2 }{4\pi}} \left[N(N-1)\ln{\frac{R}{R_0}} +{\frac{3 }{4%
}}N^2+N\ln {\frac{R}{r_c\sqrt{N}}}-{\frac{R^2}{R_0^2}}N^2\right] .
\label{G-N}
\end{eqnarray}
The condition $g=0$ yields the relation
\begin{equation}
\ln {\frac{R}{r_c}}=N\left({\frac{R^2}{R_0^2}}-{\frac{3}{4}}\right)-(N-1)\ln{%
\frac{R}{R_0}} +\ln \sqrt{N}.
\end{equation}
If the bundle nearly fills the whole container ($R_0 \approx R$, i.e., the
number $N$ of vortices is close to the equilibrium value)
\begin{equation}
\ln {\frac{R}{r_c}}\approx {\frac{N}{4}}+\ln \sqrt{N}.
\end{equation}
So the condition for equilibrium rotation of the  bundle terminating on the lateral
wall can be satisfied but not for a very large number of vortices. 
If the filling factor $R_0/R$ is small, unrealistically high
values of $\ln (R/ r_c)$ are required for equilibrium rotation of a large
number of vortices. Equilibrium rotation is impossible in the classical limit of continuous vorticity neglecting the vortex line tension $\propto \kappa \ln(r_v/r_c)$.

\section{Non-equilibrium rotation and friction} \label{fric}

The equilibrium finite vortex bundle may rotate freely only if a container
rotates with the same angular velocity or there is no friction force between
moving vortices and the container and the normal liquid rotating with it. In
the experiment this condition is not satisfied, and dynamics and structure
of the bundle require a more complicated analysis in general. But if
friction is rather weak as expected in the low-temperature limit, one may
consider the friction effect assuming that the structure of the vortex front (whorl)
is not affected seriously. 

For the state  close to the equilibrium one may use the thermodynamic approach. If the vortex front moves with  the velocity $v_f$ variation of the energy of the vortex bundle rotating with the angular velocity $\Omega$ during a short time interval $dt$ is (the energy of the front is neglected compared to that of the bundle stem)
\be
dE_b = e(\Omega) v_f dt.
    \ee{}
Displacement of the vortex front  is accompanied by variation of the bundle angular momentum $dM_z =m_z v_fdt$. The total angular momentum is conserved, and the angular momentum  $dM_z$ must be transferred from the container via mutual friction with the normal liquid moving rigidly with the container or via surface friction of vortex ends at a rough wall. This leads to energy variation of the container rotating with the angular velocity $\Omega_0$:
\be
dE_c =\Omega_0   dM_z.
      \ee{}
The total decrease of the energy due to front motion is 
\be
dE=dE_c-dE_b =-[e(\Omega)-\Omega_0m_z(\Omega)]  v_f dt.
    \ee{}
This supports the condition of the equilibrium at $\Omega=\Omega_0$ used above: the Gibbs potential density $g(\Omega)= e(\Omega)-\Omega m_z(\Omega)$ must vanish. The quantity $F_f=e(\Omega)-\Omega_0m_z(\Omega)$ can be interpreted as a force driving the vortex front in a non-equilibrium state. Since the total energy cannot change the released energy $dE$   is compensated by dissipation with the rate $\dot Q = dE/dt =  F_f v_f$. Eltsov {\em et al.} \cite{Kru} estimated the dissipation rate $\dot Q $ assuming that the bundle with $R_0 \approx R$ rotates with the same velocity as the container ($\Omega=\Omega_0$) and neglecting the quantum line-tension contribution $\propto \kappa \ln(r_v/r_c)$ to the energy. This yielded the force $F_f $ equal to the bundle kinetic energy $\pi \rho_s \Omega^2 R^4/4$ per unit length. Here we consider the state close to the equilibrium solid-body rotation at which the force $F_f=e(\Omega)-\Omega m_z(\Omega)$ vanishes. So the force is
\be
F_f=(\Omega-\Omega_0)m_z(\Omega) ={\pi \rho_s (\Omega-\Omega_0) \Omega R^4\over 2}.
    \ee{}

The vortex front can move only if there is exchange of the angular momentum between the bundle and the container.
We assume the phenomenological linear relation
between the friction torque  $t =\gamma (\Omega-\Omega_0)$ on a vortex  and the  relative
angular velocity  $\Omega-\Omega_0$.  Then the balance equation
for the total angular momentum of the liquid is
\begin{equation}
{dM_z\over dt}=m_z v_f= N t=\gamma N (\Omega-\Omega_0).
\end{equation}
For the vortex bundle with $R_0 \approx R$ the angular momentum
per unit length is $m_z= \rho_s N\kappa R^2/4$, and the front velocity
\begin{equation}
v_f =\frac{4 \gamma (\Omega-\Omega_0) }{\rho_s \kappa R^2}
\end{equation}
does not depend on the number of vortices in the bundle at given $\Omega-\Omega_0 $.

\section{Discussion and conclusions} \label{concl}

The paper addressed the question whether the vortex bundle terminating on
the lateral wall can rotate as a solid body and what is the
angular velocity of such ``eigenrotation''. There are two conditions for
existence of eigenrotation: (i) The whole vortex bundle including its stem
and the whorl is in an equilibrium solid-body rotation with the same speed;
(ii) The Gibbs potential per unit length of the stem is equal to the zero
Gibbs potential of the vortex-free state above the bundle. The latter
condition eventually determines the angular velocity of the eigenrotation
and was used in the past for estimation of the rotation of the vortex front
by Eltsov \emph{et al.}\cite{Twist,Kru} but without paying attention to the
first condition of equilibrium solid-body rotation of the whole bundle.
Indeed, using Eq.~(\ref{G0}) for linear density $g$ of the Gibbs potential
with the filling factor $R_0/R \approx 1$ and the logarithm term $\propto
\ln(r_v /r_c)$ neglected the condition $g=0$ yields that $\Omega_0 =\Omega /2
$. So in the coordinate frame rotating with this angular velocity the Gibbs potentials
above and below the vortex front are equal, and on the basis of it Eltsov
\emph{et al.} concluded that the front must rotate with the speed, which is
half of the rotation speed of the container. This rotation has nothing to do
with the equilibrium rotation analyzed in the present work: The vortex front
rotates twice slower than the vortex bundle stem, which must leads to steady
twisting of the vortex bundle. Application of thermodynamic-balance
arguments to so strongly non-equilibrium state requires justification, and a more rigorous
dynamical approach of the problem is wanted. The analysis of the equilibrium
and weakly non-equilibrium vortex bundle presented in the present paper
can be considered as the first step in this direction.

\section*{ACKNOWLEDGMENTS}
 The work was  supported by the grant of the Israel Academy
of Sciences and Humanities.  One of authors (S. K. N.) thanks the Racah Institute of Physics
of the Hebrew University of Jerusalem for hospitality and support and
acknowledges a partial support by the grants N 10-08-00369 and N 10-02-00514 from the
Russian Foundation of Basic Research.

% \begin{figure}
% \end{figure}

% Surround figure environment with turnpage environment for landscape
% figure
% \begin{figure}
% \end{figure}

%\appendix
%\section{}

%\begin{acknowledgments}
%\end{acknowledgments}


\end{document}